\documentclass[prl,twocolumn,showpacs,amsmath]{revtex4}
\usepackage{graphicx}

\newcommand{\Journal}[4]{#1 \textbf{#2}, #3 (#4)}

\newcommand{\bise}{Bi$_2$Se$_3$}

\begin{document}

\title{Scanning Tunneling Microscopy of Defect States in the Semiconductor Bi$_2$Se$_3$}

\author{S. Urazhdin}
\author{D. Bilc}
\author{S. H. Tessmer}
\author {S. D. Mahanti}
\affiliation{Department of Physics and Astronomy, Michigan State
University, East Lansing, MI 48824}
\author{Theodora Kyratsi}
\author{M. G. Kanatzidis}
\affiliation{Department of Chemistry, Michigan State University,
East Lansing, MI 48824}


\begin{abstract}
Scanning tunneling spectroscopy images of Bi$_2$Se$_3$ doped with
excess Bi reveal electronic defect states with a striking shape
resembling clover leaves. With a simple tight-binding model
we show that the geometry of the defect states in Bi$_2$Se$_3$ can be directly
related to the position of the originating impurities. Only the Bi defects at the
Se sites five atomic layers below the surface are experimentally observed. We show
that this effect can be explained by the interplay of defect and surface electronic structure.

\end{abstract}

\maketitle

Understanding the electronic properties of defects and the ability to control them
will be crucial  for the performance of the future microelectronic devices~\cite{kane}.
Scanning Tunneling Microscopy (STM) represents a unique tool for the studies of
defects as it combines atomic scale resolution with local
spectroscopic capability. However, STM observation and analysis of
defect states in semiconductors are complicated by surface effects such as in-gap surface states and 
reconstruction. These effects are avoided at the (110) surfaces of a number of III-V semiconducting
systems~\cite{gaasbands}, attracting extensive research~\cite{sigaas}--\cite{charge}. A number of point defect
types have been observed. However, positions of these defects with respect to
the surface plane could be inferred only from indirect observations. The interpretation
of such observations is complicated by the drastic effect the surface proximity
may have on the defect states~\cite{el2}.
 
Modeling STM measurements of defects in semiconductors is not straightforward: Approximation
of the STM images by maps of the local surface electronic density of states~\cite{tersoff} is justified
only if the charge relaxation rates of defect states significantly exceed the tunneling rate of electrons
between the tip and the sample~\cite{theory}. Tip-induced effects also need to be taken into 
account. These may include both local band bending~\cite{sigaas},
and charging of the  defect states by the tunneling current,
resulting in bias voltage-dependant lattice relaxation in the vicinity of the
defect atoms~\cite{charge}. Careful analysis is necessary to clearly separate these effects from the intrinsic
defect properties, and the bulk features of the observed defect states from the surface effects.

\begin{figure}
\includegraphics[scale=0.35]{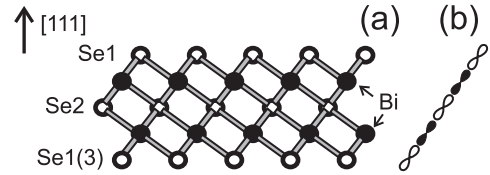}
\caption{\label{structure}
a) Structure of \bise\ showing atomic
ordering in a layer. Arrow indicates the rhombohedral [111]
layer stacking direction. In the bulk, Se1 and Se3 positions are equivalent, but we use
Se3 notation for the bottom Se atomic plane of the surface layer (with Se1 position at the surface).
(b) A schematic of the bonding into strongly $pp\sigma$ interacting chains of atoms,
5 atoms per layer. Black orbitals represent Bi; white orbitals represent Se.}
\end{figure}

In this paper we present cryogenic STM and scanning tunneling spectroscopy (STS) studies 
of the layered narrow gap semiconductor \bise, which can be viewed as a model system for the STM study
of near-surface defect states. The bonding scheme of \bise\ allows a direct determination of
the position of a subsurface defect atom with respect to the surface:
Atomic planes consisting of either Bi or Se hexagonally arranged atoms are stacked in
a close-packing fcc fashion; 5 atomic planes with atomic order Se1-Bi-Se2-Bi-Se1
(Fig.~\ref{structure}(a)) form a layer. The layers are weakly bound to each other by Se1-Se1 bonds.
Both valence and conduction bands are formed almost exclusively by the $4p$ and
$6p$ orbitals of Se and Bi respectively~\cite{bisebands}. For each atom, 
the closest neighbors from the adjacent atomic planes form
almost a regular octahedron, so the bonding can be roughly
approximated by strongly interacting $pp\sigma$ chains of atomic
$p$-orbitals (Fig.~\ref{structure}(b)), with a weaker $pp\pi$-type
interaction between adjacent chains. A substitutional defect is
therefore likely to produce a perturbation in the electronic local
density of states (DOS) predominantly along the three $pp\sigma$ chains
passing through the defect atom.  Hence, the defect state should be observed at the surface 
as three spots of modified DOS around the atoms terminating these chains at the surface.

We have performed STM and STS measurements using a custom built
low-temperature microscope with direct immersion in liquid He-4~\cite{tessmer}.
Stoichiometric \bise\ single crystal samples, as well samples doped with 2--5\% excess Bi or Se
were grown by a directional solidification technique. The stoichiometric as-grown \bise\  samples are
n-type with carrier concentration of about 10$^{19}$~cm$^{-3}$.
Doping samples with excess Bi introduces substitutional Bi defects at the 
Se sites (Bi$_{Se}$ antisites), which are shallow acceptors.
However, because of the low solubility of Bi in \bise~\cite{lovett}, the 
Bi-doped samples are n-type due to the compensating 
defects. Doping samples with excess Se introduces shallow donor-type Se
substitutional defects at the Bi sites (Se$_{Bi}$ antisites).
The samples are sufficiently inert to obtain atomic resolution in air.
However, to minimize the surface contamination, 
in the experiments reported here, the samples were cleaved {\it in situ} or in a
glove box directly attached to the STM setup in ultrapure He gas prior
to transfer to the STM with subsequent cooling to T=4.2~K.
\begin{figure}
\includegraphics[scale=0.90]{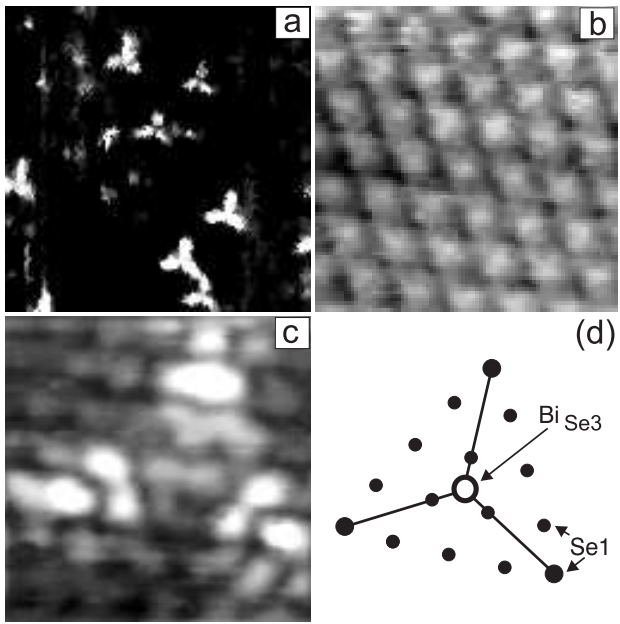}
\caption{\label{clovers}(a) A $30\times 30$~nm differential CITS map of a Bi-doped sample.
(b),(c) $3.5\times 3.5$~nm topographic maps encompassing one of the defect features.
Sample bias voltages are $-0.3$~V (b), and $-0.6$~V (c). (d) Schematic of atomic positions in (c).}
\end{figure}

To map out the defect states, we have performed differential current imaging tunneling
spectroscopy (CITS) measurements~\cite{strosciobook}. The CITS maps were acquired by
fixing the tip at each point during topographic imaging and measuring the differential
conductance at various bias voltages with the feedback loop disabled.
Only Bi-doped samples exhibited an appreciable density of observable defect states, as shown in
Fig.~\ref{clovers}(a) for a sample doped with 5\% excess Bi (Bi$_2$Se$_{2.85}$).
The map was acquired with sample bias V$_b$=$-0.2$~V in the feedback mode 
(with tunneling current set to $50$~pA), and V$_b$=$-0.45$~V for the conductivity measurement. 
The defects appear as regular clover-shaped bright features, indicating areas with locally enhanced
conductance at V$_b$=$-0.45$~V. Irregular spots in this image (mostly in the upper left corner) resulted
from topographic defects. The data discussed below were obtained on a more weakly doped Bi$_2$Se$_{2.95}$ sample,
where the defect density was reduced. 

Fig.~\ref{clovers}(b),(c) present topographic maps of the sample area encompassing an isolated 
clover-shaped defect state. Topographic image Fig.~\ref{clovers}(b), acquired at a sample bias voltage
V$_b$=$-0.3$~V, shows a periodic atomic structure, indicating no significant structural variation
associated with the defect. The height of the atomic corrugations in Fig.~\ref{clovers}(b) is about $30$~pm.
Fig.~\ref{clovers}(c) shows a topographic image of the same area acquired at V$_b$=$-0.6$~V, where the largest
corrugations, locally enhanced by tunneling through the defect state, are about $100$~pm high.
The highest amplitude of the defect state correlates with positions of
three surface Se atoms (marked with larger black circles in schematic Fig.~\ref{clovers}(d)),
forming a regular triangle. These atoms terminate three $pp\sigma$-bonded chains passing
through the Se1 site five atomic layers below the surface, for
which we also use notation Se3. Since the observed defects appear 
only in Bi-doped samples and they originate from Se sites, we 
attribute them to the Bi$_{Se3}$ antisites.

\begin{figure}
\includegraphics[scale=0.52]{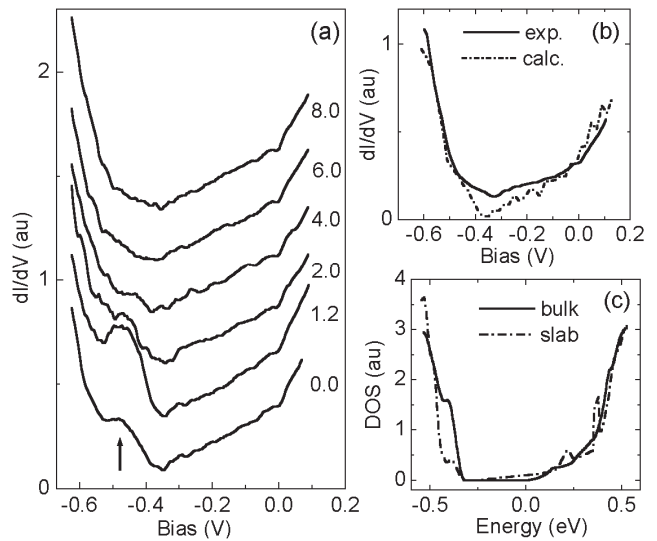}
\caption{\label{cloveriv} (a) Differential conductance spectra acquired in the vicinity of
a clover-shaped defect at various distances from the center along one of the lobes. 
Numbers on the right mark the distance from the center in nanometers; curves are offset for clarity.
(b) Differential conductance of stoichiometric \bise\ vs. calculation performed in
a slab geometry as explained in the text. (c) Calculated total
bulk near-gap density of states (DOS) {\it vs.} total DOS in the slab geometry.}
\end{figure}

Fig.~\ref{cloveriv}(a) presents a series of differential
conductance spectra acquired in the vicinity of an isolated defect.
The spectra were obtained by numerical differentiation of 60 I-V
curves with setpoint parameters V$_b$=$-0.3$~V, I=$0.8$~nA.
At the measurement temperature of 4.2~K, thermal broadening
is negligible on the displayed bias voltage scale.
The defect state appears as a broad resonance (indicated by an arrow)
in the energy range where the differential conductance is suppressed away from the defect.
The established semiconducting gap value is about $0.3$~eV~\cite{bisebands},~\cite{lovett}, therefore
the defect levels appear \emph{inside} the valence band. 
Theoretical modeling is necessary to understand this spectroscopic feature, as well as why only Bi$_{Se}$
antisites five atomic planes below the surface are observed.

First, we performed {\it ab initio} calculations in the full potential relativistic
LAPW formalism~\cite{lapw} within LDA approximation. To model the surface, a supercell geometry was used,
with distance between slabs (consisting of 15 atomic planes, or 3 layers, each)
increased by $0.5$ to $1.5$~nm as compared to the bulk crystal structure. The calculated band structure
did not exhibit significant variation for the slab separation larger than $0.4$--$-0.5$~nm, therefore we
found the slab separation of $0.7$~nm, used for the calculations presented below, sufficient for modeling the
surface. The differential conductance spectra were approximated by the local DOS in the center of the gap
between the slabs~\cite{tersoff}. The calculation presented in Fig.~\ref{cloveriv}(b) was performed for
a position above a Se1 atom, although we found the variation of the calculated spectra with position
respective to the surface atoms to be insignificant. The calculation reproduces both the finite
conductance in the bulk semiconducting gap, and the suppressed
conductance just below the gap. In Fig.~\ref{cloveriv}(c), the total DOS calculated in the slab geometry
is compared to the calculation of the bulk DOS, which reproduces the
accepted semiconducting gap value of $0.3$~eV~\cite{LDAcomment}.
Band structure analysis indicates that the highest valence band (HVB)
states are predominantly Se1-Se1 antibonding type. As the Se1-Se1 bonds are
broken at the surface, the splitting of these states is reduced, resulting in
the observed suppression of the differential conductance in the HVB energy range (Fig.~\ref{cloveriv}(b)).
The states that appear in the gap have high dispersion along the surface. They originate from the
rehybridization of the surface Se1 valence with Bi conduction states, bringing the latter down below
the bulk conduction band minimum.

\begin{figure}
\includegraphics[scale=0.73]{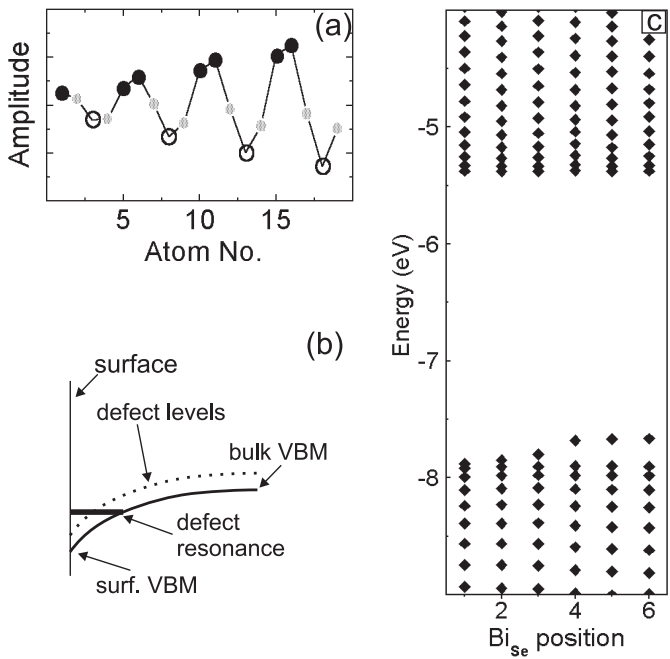}
\caption{\label{levels}  (a) Amplitudes $u_i$ (Eq.~(\ref{tb})) in the VBM state for a
16-unit chain plotted as a function of atomic position along the chain. Only the first 20 amplitudes are shown. 
Solid black circles ---Se1 positions, gray circles --- Bi positions, open circles --- Se2 positions. 
(b) Schematic of a qualitative semiclassical picture of the near-surface defect resonance formation.
(c) Calculated near-gap levels of a 16-unit chain plotted as a function of the position of Bi$_{Se}$ antisite.
The 1st position is at the surface.}
\end{figure}

{\it ab initio} calculations of a single defect state are complicated by
the large cluster or supercell size necessary to model the
impurity states without introducing artificial interaction between defects.
Instead we use a LCAO approximation as a simple
model of the system~\cite{bisebands}. This model presents just a
qualitative argument and is not capable of reproducing the detailed electronic structure or the
semiconducting gap value. However, it gives a surprisingly good qualitative
agreement with the experimental observations and first principles calculations.
In a tight binding formalism

\begin{equation} \label{tb} H\psi=H\sum_i u_i \phi_i=\sum_i E_i u_i \phi_i +
\sum_{i\ne j} V_{ij} u_i \phi_j, 
\end{equation}
where $\phi_i$ are atomic wave functions and $V_{ij}$ are off-diagonal matrix elements of $H$.
We approximate $E_i$ by the atomic term values~\cite{harrison} of Bi and Se, and
take into account only $pp\sigma$ interaction between the
closest neighbors, as shown in Fig.~\ref{structure}(b). Thus the problem is reduced to a system of
noninteracting one-dimensional chains, with three matrix elements
 $V_1$,$V_2$, and $V_3$, corresponding to Se1-Bi, Se2-Bi,
and Se1-Se1 $pp\sigma$ bonds. Consider first a
5-atom chain Se1-Bi-Se2-Bi-Se3 (we also call it a unit), representing a single layer in our
approximation. Here Se3 position is equivalent to Se1. The highest energy filled state is nonbonding

\begin{equation}
\label{nonbonding}
\psi_{0}=\frac{1}{\sqrt{2+(V_1/V_2)^2}}(\phi_{Se1}+
\phi_{Se3}-V_1/V_2\phi_{Se2}).
\end{equation}
We model the bulk by a long chain of Se1-Bi-Se2-Bi-Se1 units, with
the interaction between layers expressed by $V_3$.
Due to the Se1-Se1 interaction, the nonbonding level is split,
and the HVB states become antibonding in the sense of Se1-Se1 bond character.
The splitting is large, because the state  Eq.(\ref{nonbonding}) has a significant
weight on the interlayer Se1 atoms. As the interlayer bonds are broken at the surface,
the HVB states of a chain decay at the surface (the end of the chain), as shown in Fig.~\ref{levels}(a),
where the antibonding character of the valence band maximum (VBM) state can also be seen.
In semiclassical terms, as illustrated in Fig.~\ref{levels}(b), the
surface gap is larger than the bulk value. This effect is in agreement with the more accurate
first principles calculations and spectroscopic measurements, Fig.~\ref{cloveriv}(b), (c).

To highlight the importance of these surface effects for the observation of
the defect states, in Fig.~\ref{levels}(c) we plot the calculated
dependence of the near-gap energy levels on the position of the
Bi$_{Se}$ antisite. The second layer (positions
4--6) is only weakly affected by the proximity of the surface, so
the defect level is split from the VBM and its energy is only
weakly dependant on the position. As the antisite position
approaches the surface (positions 1--3), the defect level energy
is reduced as the surface gap opens up, so that the defect state merges with the bulk
valence states, forming a resonance \emph{inside} the bulk valence band.
This behavior is supported by the semiclassical picture shown in Fig.~\ref{levels}(b).
Only the Bi$_{Se3}$ state is observed in the experiment, because, as our model
suggests, Bi$_{Se1}$ and Bi$_{Se2}$ states are so much lowered in energy
by the proximity of the surface that they form small amplitude broad resonances in the valence band.
STM images do not exhibit defect features associated with Bi$_{Se}$ antisite in the second layer.
Our model suggests, that they form bound states in the bulk gap, which cannot sustain STM current.
Surface effects thus provide a mechanism for the charge relaxation of near-surface defect states
through the bulk valence band.

Similar surface effects should be observable in other semiconductors, e.g.
at (110) surface of GaAs, where the valence band is, like in \bise, suppressed at the
surface~\cite{gaasbands}. As a result, in-gap impurity states may become resonances in the valence band, if the 
originating impurities are sufficiently close to the surface~\cite{el2}. This suggests an alternative explanation for
some of the published results~\cite{sigaas2}. It may also be possible to induce the resonant behavior of
near-surface defects by careful control of the surface band bending with doping and/or external field.

Resonances induced by near-surface defects can be contrasted to the bulk-like in-gap states.
The origin of the spectroscopic broadening of the latter~\cite{feenstra}, and the mechanisms of their
charge relaxation, allowing their observation with STM, need further theoretical and experimental
studies~\cite{theory}. Variable temperature studies of the influence of the local defect distribution
on the spectroscopic features of defect states may provide insight into these issues, and \bise\
represents a convenient model system for such studies.

In summary, we have observed clover-shaped defect states in \bise\ doped with
excess Bi, which appear as resonances in the high valence band,  and can be
attributed to Bi$_{Se}$ antisites in fifth atomic layer from the surface. In the analysis
of these defect states, we have demonstrated the importance of the surface effects for 
the electronic structure of near-surface defects. While Bi$_{Se}$ defects in the bulk \bise\
form shallow acceptor levels, the near-surface defects produce resonances in the
energy range of the bulk valence states, which are suppressed at the surface. We suggest that
similar surface effects are likely to be observable in other semiconducting systems.

We thank J. Nogami, Norman O. Birge, M. I. Dykman, T. Hogan, and S. Lal for helpful discussions.
This work was supported in part by NSF (DMR-0075230) and ONR/DARPA
(N00014-01-1-0728). SHT acknowledges support of the Alfred P. Sloan Foundation.

\end{document}